\documentclass[twocolumn]{aastex61}

\newcommand{\RE}{R$_{\Earth}$}
\newcommand{\ME}{M$_{\Earth}$}

\shorttitle{Densities of Planets in Multiple Stellar Systems}
\shortauthors{Furlan \& Howell}
\submitjournal{AJ; accepted on June 21, 2017}

\begin{document}

\title{The Densities of Planets in Multiple Stellar Systems}

\author[0000-0001-9800-6248]{E. Furlan}
\affiliation{IPAC, Mail Code 314-6, Caltech, 1200 E. California Blvd., Pasadena, 
CA 91125, USA; furlan@ipac.caltech.edu}

\author{S. B. Howell}
\affiliation{NASA Ames Research Center, Moffett Field, CA 94035, USA}

\begin{abstract}
We analyze the effect of companion stars on the bulk density of 29 planets
orbiting 15 stars in the {\it Kepler} field. These stars have at least one stellar
companion within 2\arcsec, and the planets have measured masses and radii,
allowing an estimate of their bulk density. The transit dilution by the companion
star requires the planet radii to be revised upward, even if the planet orbits the
primary star; as a consequence, the planetary bulk density decreases. 
We find that, if planets orbited a faint companion star, they would be more
volatile-rich, and in several cases their densities would become unrealistically low, 
requiring large, inflated atmospheres or unusually large mass fractions in a H/He
envelope. In addition, for planets detected in radial velocity data, the primary star 
has to be the host. We can exclude 14 planets from orbiting the companion 
star; the remaining 15 planets in seven planetary systems could orbit either 
the primary or the secondary star, and for five of these planets the decrease in 
density would be substantial even if they orbited the primary, since the companion 
is of almost equal brightness as the primary. 
Substantial follow-up work is required in order to accurately determine the radii 
of transiting planets. Of particular interest are small, rocky planets that may be 
habitable; a lower mean density might imply a more volatile-rich composition. 
Reliable radii, masses, and thus bulk densities will allow us to identify which 
small planets are truly Earth-like.
\end{abstract}


\section{INTRODUCTION}
\label{intro}

With more than 3000 exoplanets known to date, most of them discovered by the 
{\it Kepler} mission \citep{borucki10} and increasing numbers by its successor 
{\it K2} \citep{howell14}, it has become clear that planetary systems
vary widely in their properties and that our Solar System might be in a unique
configuration. Besides the number of planets around a given star and their
orbital spacing, a fundamental quantity is a planet's density. The bulk 
density of a planet gives us clues as to its composition \citep[e.g.,][]{fortney07,
seager07,rogers11,rogers15,zeng16}: a higher density is indicative of a rocky 
interior, while a low density suggests a planet surrounded by a substantial 
atmosphere. Of particular interest are rocky planets with liquid water on their 
surface and an atmosphere, which, if at a suitable distance from their star, 
might be able to support life as we know it.

In order to determine a planet's mean density, its mass and radius have to be known.
The {\it Kepler} mission discovered planets by the transit method, which measures
the dimming of the stellar light as the planet passes in front of its star. The observed
transit depth yields the radius of the planet, assuming the stellar radius is known.
The mass is typically determined from radial velocity (RV) follow-up measurements of
the planet \citep[e.g.,][]{marcy14}; in some cases of multiple planetary systems, 
transit-timing variations (TTVs) can be used to determine planetary masses \citep[e.g.,][]
{hadden14}. Uncertainties in the determination of the planet's radius and mass 
propagate to uncertainties in the planet's density.

Besides the usual measurement uncertainties, one factor can affect the reliable
determination of a planet's radius: the presence of one or more stellar companions.
The transit method derives the planet's radius from the transit depth, which is the
difference of the out-of-transit and in-transit flux relative to the out-of-transit flux. 
A stellar companion dilutes the transit, making it appear shallower, and thus we 
infer a smaller planetary radius. Therefore, the presence of close companions leads 
to an underestimate of planetary radii. These companions are not necessarily
bound to the primary star; studies of {\it Kepler} stars have shown that most companions 
within 1\arcsec\ are bound, while this applies to only $\sim$ 50\% of companions at 
2\arcsec\ \citep{horch14,hirsch17}. However, even a close background star will dilute 
the transit and require a revision of the derived planet radius.

When planetary radii are underestimated, their density is overestimated, which
is an issue of particular importance for small, rocky, potentially habitable planets. With a
close companion star present, the radius of such a ``small'' planet would have 
to be revised upward, possibly requiring a substantial gaseous envelope to 
explain the resulting lower bulk density. Recently, seven Earth-sized planets
were discovered transiting the nearby star TRAPPIST-1 \citep{gillon16,gillon17};
their densities suggest a rocky composition with a certain fraction of volatiles 
\citep{gillon17}. \citet{howell16} carried out speckle imaging of TRAPPIST-1
and were able to exclude a companion star or brown dwarf from 0.32 to 14.5
au from the star; their results complemented the RV measurements from 
\citet{barnes14}, which ruled out stellar companions within about 0.15 au.
Thus, follow-up observations established that the radii of the TRAPPIST-1 planets 
derived from transits are correct.

For the {\it Kepler} mission, a substantial imaging and spectroscopic follow-up 
observation program was carried out (for a summary, see \citealt{furlan17a} 
and references therein; Furlan et al. 2017b, in preparation). The aim of the 
imaging program was to detect companion stars to planet host stars, while 
the main goal of the spectroscopic program was to refine stellar parameters.
RV measurements (which require high spectral resolution) are mainly used 
to determine planet masses, but they can also reveal close companion stars 
\citep{kolbl15}. However, only a certain range of parameter space can be probed
by spectroscopy; companions that are too faint, too far, or too similar to the 
primary star cannot be detected. \citet{teske15} showed that the RV detections 
can be very uncertain; beyond about 0.02\arcsec, high-resolution imaging yields 
more reliable and complete information on stellar companions. From the compilation 
of high-resolution and seeing-limited imaging of KOI host stars in \citet{furlan17a}, 
we find that about 6\% (11\%) of the detected companions lie within 0.5\arcsec\ 
(1.0\arcsec) from their primary stars and have median $\Delta$m values of 0.9 (1.5) 
in the $K$-band and 1.0 (1.3) in the $i$-band.

From the solar neighborhood, we know that about 44\% of solar-type stars
have a bound companion within $\sim$ 10,000 au, with most companions 
at separations between a few and a few hundred au \citep{raghavan10}.
The multiplicity of stars in the {\it Kepler} field, which lie at distances 
up to a few kpc (the median distance is 840 pc; \citealt{mathur17}) has not yet 
been well-established. 
\citet{horch14} carried out simulations of the {\it Kepler} field using a companion 
star fraction of 40\%--50\% and the distribution of binaries in the solar neighborhood 
\citep{duquennoy91,raghavan10}, and they were able to reproduce their observed 
companion star fractions from speckle observations. Their results implied that about 
half of {\it Kepler} stars have companions, even though not all of them can be 
detected. However, several recent studies found lower stellar multiplicity rates for 
host stars of KOI planets, especially at projected separations less than a few tens 
up to a few hundred au \citep{wang14a,wang14b,wang15a,wang15b,kraus16}.
On the other hand, due to detection and sensitivity limits, some parts of the binary
parameter space, e.g.\ companions at separations $\lesssim$ 10 au (accessible only 
via RV measurements) or companions with $\Delta$m $\gtrsim$ 3 at $\lesssim$ 
20 au (in high-resolution images) have not yet been fully explored.

The detectability of stellar companions does not only depend on their
projected separations from the primary star, but also their relative brightness.
\citet{raghavan10} found that the mass-ratio distribution for stars in multiple 
systems is mostly flat, with a deficit at low values ($\lesssim$ 0.2), but a sharp 
increase in the number of companions with mass ratios close to unity. From 
the data presented in \citet{raghavan10}, we deduce that the fraction of about 
equal-mass systems (mass ratio $>$ 0.9) is 17$\pm$3\%; this fraction increases 
to 27$\pm$5\%, 30$\pm$6\%, and 38$\pm$10\% for stars with about equal-mass 
companions within 100, 50, and 10 au, respectively. Thus, we can infer that about 
15\% of stars (at least in the solar neighborhood, perhaps also in the {\it Kepler}
field) have such bright, close companions; it is this type of companions that have 
the strongest effect on derived planet radii if planets are assumed to orbit their 
primary star. Equal-brightness binaries increase the planet radius (derived under 
the assumption that the star is single) the most, namely by a factor of 1.4. Planets
that orbit a star with a fainter companion typically have radii overestimated by a 
few percent \citep{furlan17a}.

A scenario rarely considered in the literature is the possibility that a planet 
could orbit a fainter companion star. In this case its radius would need a correction
by a factor of a few \citep{furlan17a}. It is necessary to assess each system to 
determine which star the planet likely orbits, but in some cases, the companion star 
can be excluded as being the host star based on the lack of significant centroid shifts 
\citep[e.g.,][]{latham10,bryson13} or on the color of the companion star 
\citep[e.g.,][]{howell12,hirsch17}. In other cases, more thorough follow-up work, 
especially a statistical analysis of the available data, is needed to determine the 
actual host star and thus an accurate planet radius \citep[e.g.,][]{barclay15}.
We note that in cases of very close stellar companions ($\lesssim$ a few au),
planets might actually orbit both stars. In fact, there are planets known to orbit 
eclipsing binary stars in the {\it Kepler} field \citep[e.g.,][]{doyle11,welsh12,
orosz12,schwamb13,kostov16}. Since the radii of eclipsing binary stars can be 
measured  quite accurately, the radii of planets orbiting them are fairly reliable, too.

In \citet{furlan17a}, we calculated planet radius correction factors for all those 
{\it Kepler} planet host stars with a stellar companion within 4\arcsec. We
assumed companion stars to be bound to the primary stars and thus at the same 
distance from Earth, so properties such as their stellar radius could be estimated.
Our results agreed with those from \citet{ciardi15}, who used the multiplicity fraction 
and mass ratio distribution from \citet{raghavan10} and estimated that, on average, 
the radii of {\it Kepler} planets are underestimated by a factor of 1.5.

In this work, we use the results presented in \citet{furlan17a} and apply them to 
{\it Kepler} planets whose masses have been determined in addition to the radii
derived from the transit observations. We estimate the change in radius and thus 
density for the planets and discuss the implications for the planets' composition. 
We present our sample in Section \ref{sample}, our results in Section \ref{res}, and
our discussion in Section \ref{discuss}; Section \ref{conclude} contains our conclusions.

\section{SAMPLE}
\label{sample}

In \citet{furlan17a}, we combined measurements of detected companions within
4\arcsec\ (one {\it Kepler} pixel) of host stars of {\it Kepler} Objects of Interest (KOIs) 
and created a catalog of 2297 companions around 1903 primary stars. The KOIs 
can be either planet candidates or false positives; only follow-up observations (radial
velocity measurements, high-resolution imaging) can confirm a planet candidate as 
an actual planet, but planets have also been validated by analyzing observational 
results with statistical methods \citep[see, e.g.,][]{rowe14,morton16}. 
Here we only select {\it Kepler} stars which are hosts to confirmed planets and 
have one or more companions within 2\arcsec\ listed in \citet{furlan17a}. 
Companions at these projected separations are more likely to be bound 
\citep[see][]{horch14,hirsch17} and are also unlikely to be detected by the 
{\it Kepler} photometric centroid shift analysis \citep{bryson13}; also, 
none of these companions are listed in the Kepler Input Catalog (KIC).
A close companion, even if unbound, will dilute the transit depth and thus 
affect the derived planet radius. Moreover, we limit our sample to confirmed 
{\it Kepler} planets with measured masses (including upper
limits) and radii, which allows us to infer the bulk density of the planets.
Additionally, we exclude those planets from further analysis for which 
no correction to the planet radius is needed, as detailed below.

Table \ref{planet_overview} lists all confirmed {\it Kepler} planets with
masses, radii, and at least companion star within 2\arcsec\ from the compilation
\citet{furlan17a}. This sample amounts to 50 planets orbiting 26 stars. We 
adopted planetary mass and radius measurements from the literature (as 
collected by the NASA Exoplanet Archive\footnote{http://exoplanetarchive.ipac.caltech.edu}). 
When more than one measurement was available, we calculated a weighted 
average using the inverse of the uncertainty as weights.
The column ``blend flag'' in Table \ref{planet_overview} indicates whether
authors already included the effect of nearby companion stars in their analysis 
of the {\it Kepler} light curves. The radii of Kepler-1 b, Kepler-5 b, Kepler-7 b, 
Kepler-13 b, Kepler-14 b, Kepler-64 b, and Kepler-432 b are already corrected 
for flux dilution by the nearby companion star. In most cases, this flux dilution is 
just a few tenths to a few percent and therefore the change in the resulting planet 
radius small \citep[e.g.,][]{esteves15}. The largest corrections to the transit depth 
(and thus planet radii) were applied for Kepler-13 b, Kepler-14 b, and Kepler-64 b  
\citep{szabo11,shporer14,esteves15, buchhave11,southworth12,schwamb13}.
We note that in general, even when the effect of the companion was 
included in the derivation of planet radii, usually only the case of planets orbiting 
their primary star was considered. The planet radius would change substantially 
if the planet orbited a fainter companion star.

The column ``mass flag'' in Table \ref{planet_overview} identifies whether the
mass of a planet was determined from RV measurements, TTVs, or a light curve
model (in some cases a combined model to multiple data sets; e.g., \citealt{schwamb13}).
In cases where the planet mass was derived via RV measurements, it is clear that 
planets are orbiting the primary star (whose RV variations have been measured).
Therefore, the companion stars in the Kepler-1, Kepler-5, Kepler-7, Kepler-10,
Kepler-14, Kepler-21, Kepler-64, Kepler-74, Kepler-97, Kepler-106, Kepler-424,
Kepler-432, and Kepler-448 systems cannot be the planet host stars. For Kepler-100,
the situation is less clear, since planets c and d were not detected in the RV data,
and planet b only had a tentative detection \citep{marcy14}. So, we keep the 
possibility open that the Kepler-100 planets could orbit the companion star. Finally, 
based on centroid analysis of {\it Kepler} data, the primary stars in the Kepler-11 
and Kepler-13 systems were determined to be the ones transited by the planets 
\citep{lissauer11, szabo11}.

For this work, we do not further consider those planets for which the 
companion star was excluded to be the planet host and its flux dilution has already
been accounted for in the derived planet parameters. In addition, we also remove 
from our sample the Kepler-10, Kepler-11, Kepler-21, Kepler-106, and Kepler-424 
systems, since the primary stars were found to be the planet hosts, and the flux 
dilution by the companion, while not corrected for, is very minute ($\lesssim$ 0.5\%).
The final sample we analyze in this work consists of 29 planets orbiting 15 stars
(see Table \ref{planet_densities}). As with the planets' masses and radii, we adopted
density measurements from the literature. In some cases, for a given planet only 
a mass ($M$) and radius ($R$) were published, but not the density; in those cases 
we carried out a simple calculation of the mean density (${\rho}=M/(\frac{4}{3}{\pi}R^3)$, 
$\Delta{\rho}/{\rho} = \sqrt{\left(\frac{\Delta M}{M}\right)^2+9\left(\frac{\Delta R}{R}\right)^2}$).
For published densities, we adopted the reported measurements and their
uncertainties. When more than one density value was available for a given
planet, we calculated a weighted average as we did for masses and radii.
Planets with just an upper limit for their mass only have an upper limit for their 
density.
Some planets have unrealistically high densities, both in published values and
from our simple calculation. The likely reason is an overestimate of their masses;
in several cases the masses were determined from TTVs, and a substantial 
underestimate of the orbital eccentricities leads to an overestimate of the planetary 
masses \citep[there is a degeneracy between these two parameters; see][]{hadden14}. 
In other cases the masses determined from radial velocities are very uncertain 
\citep[e.g.,][]{marcy14}, resulting in large uncertainties in the derived bulk densities.

Also listed in Table \ref{planet_overview} are the planet radius correction factors
(PRCF) from \citet{furlan17a}; since they only depend on stellar parameters, each planet
in a multi-planet systems has the same radius correction factor. Multiplying the planet 
radius by these factors yields the actual planet radius. There are two sets of factors: 
one assuming that planets orbit their primary star (``primary'' factor hereafter), and one 
assuming planets orbit the brightest companion star (under the assumption that it is bound 
to the primary star; ``secondary'' factor hereafter). The former is close to 1.0 in most cases; 
it is largest for Kepler-326 and Kepler-84, which each have a nearby companion 
of almost equal brightness (at 0.05\arcsec\ with ${\Delta}K$=0.03 for Kepler-326; 
at 0.2\arcsec\ with ${\Delta}$m $\sim$ 0.9 at 0.55 $\mu$m for Kepler-84; 
\citealt{kraus16,gilliland15}). The radii of the planets in these two systems
were derived from stellar radii and planet-to-star size ratios as reported in the literature, 
which do not seem to take into account the presence of the bright, nearby companions 
\citep{hadden14}. No primary correction factor is listed for those planets for
which the flux dilution by the companion has already been accounted for when the
planet radius was derived.

For the secondary planet radius correction factors, there is a limit on how large
they can be: the planet can only become as large as the companion star
(thus obscuring 100\% of the companion star during transit), which would also imply
that it is likely not a planet, but a star. In these cases (Kepler-5, Kepler-106, 
Kepler-145, Kepler-424), the planet host stars do not have a secondary correction 
factor; moreover, the companion star is so faint that the primary correction factor 
is very small, less than 1\%. The secondary factor is also not listed for those 
planets determined to orbit the primary star.

The planet radius correction factors can be converted to planet density
correcting factors (PDCF), as  $\mathrm{PDCF} = \mathrm{PRCF}^{-3}$. 
These factors are listed in Table \ref{planet_densities}, with one set assuming 
that planets orbit the primary star and one set assuming planets orbit the brightest 
companion star. There is no secondary PDCF if planets were determined
to orbit the primary star, which includes those systems in which companion stars 
could be excluded as being the planet hosts due to the measured transit depth 
(see above). We used the calculated density correction factors to correct the 
planet bulk densities; these corrected densities are also listed in Table 
\ref{planet_densities}.

\startlongtable
\begin{longrotatetable}
\begin{deluxetable*}{lllccccccc}\scriptsize
\movetabledown=1.7in
%\tabletypesize{footnotesize}
\tablewidth{0pt}
\tablecaption{Masses, Radii, and Planet Radius Correction Factors of Confirmed {\it Kepler} 
Planets Orbiting Stars with Stellar Companions at $\leq$ 2\arcsec
\label{planet_overview}}
\tablehead{
\colhead{Planet Name} & \colhead{KOI} & \colhead{KICID} & \colhead{Mass [M$_J$]} & 
\colhead{Radius [R$_J$]} & \colhead{Mass Flag} & \colhead{Blend Flag} &
\colhead{PRCF$_{p}$} & \colhead{PRCF$_{s}$} & \colhead{Ref.} \\
\colhead{(1)} & \colhead{(2)} & \colhead{(3)} & \colhead{(4)} & \colhead{(5)} & \colhead{(6)} & 
\colhead{(7)} & \colhead{(8)} & \colhead{(9)} & \colhead{(10)}} 
\startdata
  Kepler-1b &    1 & 11446443 &  1.2232$\pm$0.018 &  1.213$\pm$0.011 & R,M & 1 & \nodata &  \nodata & 1,6,8,11,15,16,24,28,29,31,32,33 \\
  Kepler-5b &   18 &  8191672 &  2.0818$\pm$0.033 &  1.339$\pm$0.023 & R,M & 1 & \nodata &  \nodata & 11,19,29 \\
  Kepler-7b &   97 &  5780885 &  0.4367$\pm$0.024 &  1.604$\pm$0.015 & R,M &  1 & \nodata &  \nodata & 11,17,29,30 \\
 Kepler-10b &   72 & 11904151 &  0.0126$\pm$0.002 &  0.130$\pm$0.001 & R & 0  & 1.000 &  \nodata & 2,9,11,12 \\
 Kepler-10c &   72 & 11904151 &  0.0540$\pm$0.006 &  0.210$\pm$0.006 & R & 0  & 1.000 &  \nodata & 9 \\
 Kepler-11b &  157 &  6541920 &  0.0105$\pm$0.003 &  0.161$\pm$0.004 & T & 0  & 1.003 &  \nodata & 13,20,21 \\
 Kepler-11c &  157 &  6541920 &  0.0187$\pm$0.006 &  0.257$\pm$0.005 & T  & 0  & 1.003 &  \nodata & 13,20,21 \\
 Kepler-11d &  157 &  6541920 &  0.0215$\pm$0.003 &  0.279$\pm$0.006 & T  & 0  & 1.003 &  \nodata & 13,20,21 \\
 Kepler-11e &  157 &  6541920 &  0.0249$\pm$0.005 &  0.374$\pm$0.008 & T  & 0 & 1.003 &  \nodata & 13,20,21 \\
 Kepler-11f &  157 &  6541920 &  0.0067$\pm$0.003 &  0.221$\pm$0.005 & T  & 0 & 1.003 &  \nodata & 13,20,21 \\
 Kepler-11g &  157 &  6541920 & $<$ 0.0790 &  0.297$\pm$0.006 & T & 0 & 1.003 &  \nodata & 21 \\
 Kepler-13b &   13 &  9941662 &  9.0250$\pm$0.205 &  1.461$\pm$0.026 & M & 1 & \nodata &  \nodata & 11,27 \\
 Kepler-14b &   98 & 10264660 &  8.0620$\pm$0.259 &  1.130$\pm$0.040 & R,M & 1 & \nodata &  \nodata & 5,30 \\
 Kepler-21b &  975 &  3632418 &  0.0160$\pm$0.005 &  0.146$\pm$0.001 & R & 0 & 1.002 &  \nodata & 18 \\
 Kepler-27b &  841 &  5792202 &  0.1320$\pm$0.018 &  0.522$\pm$0.024 & T & 0 & 1.014 &  3.430 & 13 \\
 Kepler-27c &  841 &  5792202 &  0.0670$\pm$0.011 &  0.640$\pm$0.029 & T  & 0 & 1.014 &  3.430 & 13 \\
 Kepler-53b &  829 &  5358241 &  0.3240$\pm$0.106 &  0.253$\pm$0.061 & T & 0 & 1.054 &  1.820 & 13 \\
 Kepler-53c &  829 &  5358241 &  0.1120$\pm$0.053 &  0.278$\pm$0.067 & T & 0 & 1.054 &  1.820 & 13 \\
 Kepler-64b & 6464 &  4862625 & $<$ 0.5310 &  0.551$\pm$0.015 & R,M & 1 & \nodata &  \nodata & 26 \\
 Kepler-74b &  200 &  6046540 &  0.6586$\pm$0.073 &  1.005$\pm$0.025 & R & 0 & 1.032 &  \nodata & 3,14 \\
 Kepler-80b &  500 &  4852528 &  0.0218$\pm$0.002 &  0.238$\pm$0.009 & T & 0 & 1.002 &  5.343 & 22 \\
 Kepler-80c &  500 &  4852528 &  0.0212$\pm$0.004 &  0.244$\pm$0.010 & T & 0 & 1.002 &  5.343 & 22 \\
 Kepler-80d &  500 &  4852528 &  0.0212$\pm$0.002 &  0.136$\pm$0.007 & T & 0 & 1.002 &  5.343 & 22 \\
 Kepler-80e &  500 &  4852528 &  0.0130$\pm$0.003 &  0.143$\pm$0.007 & T & 0 & 1.002 &  5.343 & 22 \\
 Kepler-84b & 1589 &  5301750 &  0.1260$\pm$0.038 &  0.174$\pm$0.045 & T & 0 & 1.202 &  1.387 & 13 \\
 Kepler-84c & 1589 &  5301750 &  0.0640$\pm$0.037 &  0.184$\pm$0.047 & T & 0 & 1.202 &  1.387 & 13 \\
 Kepler-92b &  285 &  6196457 &  0.2020$\pm$0.044 &  0.313$\pm$0.009 & T  & 0 & 1.002 &  3.079 & 34 \\
 Kepler-92c &  285 &  6196457 &  0.0190$\pm$0.006 &  0.232$\pm$0.007 & T & 0 & 1.002 &  3.079 & 34 \\
 Kepler-97b &  292 & 11075737 &  0.0110$\pm$0.006 &  0.132$\pm$0.012 & R & 0 & 1.029 &  \nodata & 23 \\
Kepler-100b &   41 &  6521045 &  0.0230$\pm$0.010 &  0.118$\pm$0.004 & R & 0 & 1.008 &  3.604 & 23 \\
Kepler-100c &   41 &  6521045 & $<$ 0.0222 &  0.196$\pm$0.004 & R & 0 &  1.008 & 3.604 & 23 \\
Kepler-100d &   41 &  6521045 & $<$ 0.0094 &  0.144$\pm$0.004 & R & 0 & 1.008 &  3.604 & 23 \\
Kepler-104b &  111 &  6678383 &  0.0620$\pm$0.043 &  0.279$\pm$0.054 & T & 0 & 1.001 &  6.059 & 13 \\
Kepler-106b &  116 &  8395660 & $<$ 0.0167 &  0.073$\pm$0.010 & R & 0 & 1.000 &  \nodata & 23 \\
Kepler-106c &  116 &  8395660 &  0.0330$\pm$0.010 &  0.223$\pm$0.029 & R & 0 & 1.000 &  \nodata & 23 \\
Kepler-106d &  116 &  8395660 & $<$ 0.0255 &  0.085$\pm$0.012 & R & 0 & 1.000 &  \nodata & 23 \\
Kepler-106e &  116 &  8395660 &  0.0350$\pm$0.018 &  0.228$\pm$0.029 & R & 0 & 1.000 &  \nodata & 23 \\
Kepler-145b &  370 &  8494142 &  0.1170$\pm$0.036 &  0.236$\pm$0.007 & T & 0 & 1.000 &  \nodata & 34 \\
Kepler-145c &  370 &  8494142 &  0.2500$\pm$0.052 &  0.385$\pm$0.011 & T & 0 & 1.000 &  \nodata & 34 \\
Kepler-203c &  658 &  6062088 &  2.3600$\pm$1.202 &  0.186$\pm$0.046 & T & 0 & 1.009 &  3.326 & 13 \\
Kepler-203d &  658 &  6062088 &  0.1070$\pm$0.340 &  0.109$\pm$0.027 & T & 0 & 1.009 &  3.326 & 13 \\
Kepler-326b & 1835 &  9471268 &  0.1400$\pm$0.127 &  0.270$\pm$0.159 & T & 0 & 1.407 &  1.421 & 13 \\
Kepler-326c & 1835 &  9471268 &  0.0550$\pm$0.041 &  0.249$\pm$0.146 & T & 0 & 1.407 &  1.421 & 13 \\
Kepler-326d & 1835 &  9471268 &  0.0220$\pm$0.023 &  0.215$\pm$0.126 & T & 0 & 1.407 &  1.421 & 13 \\
Kepler-333b & 1908 &  5706966 &  0.0890$\pm$0.083 &  0.144$\pm$0.015 & T & 0 & 1.009 &  2.970 & 13 \\
Kepler-396b & 2672 & 11253827 &  0.2380$\pm$0.027 &  0.312$\pm$0.086 & T & 0 & 1.001 &  2.733 & 34 \\
Kepler-396c & 2672 & 11253827 &  0.0560$\pm$0.007 &  0.473$\pm$0.131 & T & 0 & 1.001 &  2.733 & 34 \\
Kepler-424b &  214 & 11046458 &  1.0300$\pm$0.130 &  0.890$\pm$0.070 & R & 0 & 1.001 &  \nodata & 10 \\
Kepler-432b & 1299 & 10864656 &  5.2251$\pm$0.232 &  1.132$\pm$0.026 & R & 1  & \nodata &  \nodata & 7,25 \\
Kepler-448b &   12 &  5812701 & $<$10.0 &  1.430$\pm$0.130 & R & 0 & 1.021 &  \nodata & 4 \\
\enddata
\tablecomments{Column (1) lists the {\it Kepler} planet name, column (2)  the KOI number of the star, 
column (3) its identifier from the {\it Kepler} Input Catalog (KIC), column (4) the mass of the planet,
column (5) the radius of the planet, column (6) identifies the methods by which the mass was determined
('R' --- RV, 'T' --- TTV, 'M' --- light curve model),
column (7) indicates whether the blending by a nearby companion was already accounted for 
when the planet radius was derived in at least one of the references listed in column (10) 
(1--- yes, 0 --- no),
columns (8) and (9) list the planet radius correction factors assuming the planet orbits the primary 
or brightest secondary star, respectively, from \citet{furlan17a}, and column (10) lists the references
for planet mass and radius.}
\tablerefs{(1) \citet{barclay12}; (2) \citet{batalha11}; (3) \citet{bonomo15}; (4) \citet{bourrier15}; 
(5) \citet{buchhave11}; (6) \citet{christiansen11}; (7) \citet{ciceri15}; (8) \citet{daemgen09}; 
(9) \citet{dumusque14}; (10) \citet{endl14}; (11) \citet{esteves15}; (12) \citet{fogtmann-schulz14}; 
(13) \citet{hadden14}; (14) \citet{hebrard13}; (15) \citet{holman07}; (16) \citet{kipping11}; 
(17) \citet{latham10}; (18) \citet{lopez-morales16}; (19) \citet{koch10}; (20) \citet{lissauer11}; 
(21) \citet{lissauer13}; (22) \citet{macdonald16}; (23) \citet{marcy14}; (24) \citet{odonovan06}; 
(25) \citet{quinn15}; (26) \citet{schwamb13}; (27) \citet{shporer14}; 
(28) \citet{southworth10}; (29) \citet{southworth11}; (30) \citet{southworth12}; 
(31) \citet{sozzetti07}; (32) \citet{torres08}; (33) \citet{turner16}; (34) \citet{xie14}}
\end{deluxetable*}
\end{longrotatetable}

\startlongtable
\begin{longrotatetable}
\begin{deluxetable*}{lcccccccc}\scriptsize
\movetabledown=1.7in
%\tabletypesize{footnotesize}
\tablewidth{0pt}
\tablecaption{Bulk Densities, Planet Density Correction Factors, Orbital Periods, and
Equilibrium Temperatures of {\it Kepler} Planets Studied in this Work
\label{planet_densities}}
\tablehead{
\colhead{Planet Name} & \colhead{$\rho$ [g cm$^{-3}$]} & \colhead{PDCF$_{p}$} & 
\colhead{PDCF$_{s}$} &  \colhead{$\rho_{\mathrm{corr,p}}$ [g cm$^{-3}$]} 
& \colhead{$\rho_{\mathrm{corr,s}}$ [g cm$^{-3}$]} & \colhead{P [d]} &
\colhead{$T_{eq}$ [K]} & \colhead{Ref.} \\
\colhead{(1)} & \colhead{(2)} & \colhead{(3)} & \colhead{(4)} & \colhead{(5)} & \colhead{(6)} & 
\colhead{(7)} & \colhead{(8)} & \colhead{(9)}} 
\startdata
 Kepler-27b &      1.151$\pm$ 0.220 & 0.9597 &  0.0248 &      1.105 &      0.029 & 15.33 & 610 & 8,12,13 \\
 Kepler-27c &      0.317$\pm$ 0.068 & 0.9597 &  0.0248 &      0.304 &      0.008 & 31.33 & 481 & 8,12,13 \\
 Kepler-53b &     24.811$\pm$19.563 & 0.8540 &  0.1658 &     21.190 &      4.113 & 18.65 & 701 & 9,12,13 \\
 Kepler-53c &      6.465$\pm$ 5.573 & 0.8540 &  0.1658 &      5.521 &      1.072 & 38.56 & 550 & 9,12,13 \\
 Kepler-74b &      0.584$\pm$ 0.107 & 0.9101 &  \nodata &      0.531 &     \nodata & 7.34 & 1164 & 1,3 \\
 Kepler-80b &      1.380$\pm$ 0.205 & 0.9952 &  0.0066 &      1.373 &      0.009 & 7.05 & 546 & 4,6 \\
 Kepler-80c &      1.220$\pm$ 0.205 & 0.9952 &  0.0066 &      1.214 &      0.008 & 9.52 & 494 & 4,6 \\
 Kepler-80d &      7.040$\pm$ 1.060 & 0.9952 &  0.0066 &      7.006 &      0.046 & 3.07 & 720 & 4,6 \\
 Kepler-80e &      3.750$\pm$ 0.930 & 0.9952 &  0.0066 &      3.732 &      0.025 & 4.64 & 628 & 4,6 \\
 Kepler-84b &     29.661$\pm$24.648 & 0.5751 &  0.3745 &     17.058 &     11.109 & 8.73 & 985 & 10,12,13 \\
 Kepler-84c &     12.740$\pm$12.313 & 0.5751 &  0.3745 &      7.327 &      4.772 & 12.88 & 865 & 10,12,13 \\
 Kepler-92b &      8.169$\pm$ 1.914 & 0.9943 &  0.0343 &      8.123 &      0.280 & 13.75 & 975 & 11,12,13 \\
 Kepler-92c &      1.887$\pm$ 0.620 & 0.9943 &  0.0343 &      1.876 &      0.065 & 26.72 & 781 & 11,12,13 \\
 Kepler-97b &      5.440$\pm$ 3.480 & 0.9175 &  \nodata &      4.991 &    \nodata & 2.59 & 1328 & 12,5 \\
Kepler-100b &     14.250$\pm$ 6.330 & 0.9755 &  0.0214 &     13.901 &     0.304 & 6.89 & 1155 & 5,12 \\
Kepler-100c & $<$  3.653 & 0.9755 &  0.0214 & $<$  3.564 & $<$  0.078 & 12.82 & 939 & 5,12,13 \\
Kepler-100d & $<$  3.921 & 0.9755 &  0.0214 & $<$  3.825 & $<$  0.084 & 35.33 & 670 & 5,12,13 \\
Kepler-104b &      3.540$\pm$ 3.180 & 0.9961 &  0.0045 &      3.526 &      0.016 & 11.43 & 852 & 7,12,13 \\
Kepler-145b &     11.038$\pm$ 3.536 & 0.9997 &  \nodata  &  11.035 &      \nodata & 22.95 & 873 & 11,12,13 \\
Kepler-145c &      5.433$\pm$ 1.222 & 0.9997 &  \nodata  &  5.431 &      \nodata & 42.88 & 709 & 11,12,13 \\
Kepler-203c &    \nodata & 0.9726 &  0.0272 &  \nodata &    \nodata & 5.37 & 1096 & 7,12,13 \\
Kepler-203d &    \nodata & 0.9726 &  0.0272 &  \nodata &    \nodata & 11.33 & 855 & 7,12,13 \\
Kepler-326b &      8.821$\pm$17.488 & 0.3589 &  0.3486 &      3.166 &      3.075 & 2.25 & 1127 & 7,12,13 \\
Kepler-326c &      4.418$\pm$ 8.440 & 0.3589 &  0.3486 &      1.585 &      1.540 & 4.58 & 889 & 7,12,13 \\
Kepler-326d &      2.745$\pm$ 5.632 & 0.3589 &  0.3486 &      0.985 &      0.957 & 6.77 & 781 & 7,12,13 \\
Kepler-333b &     36.962$\pm$36.551 & 0.9729 &  0.0382 &     35.960 &      1.410 & 12.55 & 480 & 7,12,13 \\
Kepler-396b &      9.718$\pm$ 8.114 & 0.9970 &  0.0490 &      9.689 &      0.476 & 42.99 & 496 & 11,12,13 \\
Kepler-396c &      0.656$\pm$ 0.551 & 0.9970 &  0.0490 &      0.654 &      0.032 & 88.50 & 390 & 11,12,13 \\
Kepler-448b & $<$  4.241 & 0.9382 &  \nodata & $<$  3.979 & \nodata & 17.85 & 911 & 2,12,13 \\
\enddata
\tablecomments{Column (1) lists the {\it Kepler} planet name, column (2) the planet density 
(either from the literature or derived in this work; see text for details), columns (3) and (4) 
the planet density correction factors assuming the planet orbits the primary or brightest 
secondary star, respectively, columns (5) and (6) the planet densities corrected using the
factors from columns (3) and (4), respectively, column (7) the planet's orbital period, column (8)
the planet's equilibrium temperature, and column (9) the references for the planet parameters 
listed.}
\tablerefs{(1) \citet{bonomo15}; (2) \citet{bourrier15}; (3) \citet{hebrard13}; (4) \citet{macdonald16}; 
(5) \citet{marcy14}; (6) \citet{muirhead12}; (7) \citet{rowe14}; (8) \citet{steffen12}; (9) \citet{steffen13};
(10) \citet{xie13}; (11) \citet{xie14}; (12) Q1-Q17 DR25 KOI table, (13) this work}
\end{deluxetable*}
\end{longrotatetable}

\section{RESULTS}
\label{res}

\subsection{Effect of Companions on Planet Bulk Density}

\begin{figure}[!t]
\centering
\includegraphics[scale=0.49]{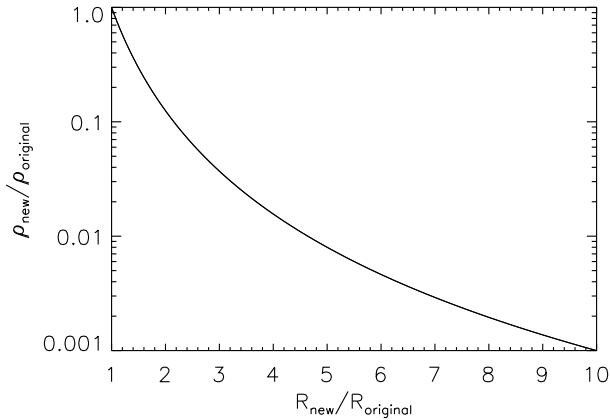}
\caption{Fractional change of planet density versus fractional change of 
planet radius. The subscript ``new'' identifies the new, corrected values,
while the subscript ``original'' stands for the originally derived parameter
value (for example, not taking into account the presence of a stellar
companion).
\label{density_radius}}
\end{figure}

\citet{ciardi15} estimated the effect of stellar companions on the derived 
planetary radii of all KOIs; they assumed that KOI host stars could be single
or in binary or triple systems, and, in the case of multiple systems, the 
planets could orbit the primary star or one of the companion stars. They
also assumed that the multiplicity of stars in the {\it Kepler} field is similar
to that of stars in the solar neighborhood, as derived by \citet{raghavan10}
and estimated by \citet{horch14}. On average, they found that planet radii 
are underestimated by a factor of 1.49.
In \citet{furlan17a} we used the compiled measurements on 1903 KOI
host stars with companions detected within 4\arcsec; the median correction
factors for planet radii assuming planets orbit the primary or brightest 
companion star were 1.01 and 2.69, respectively. A weighted average of
these correction factors yielded a median value of 1.38 if planets were assumed
more likely to orbit the primary star; if assuming that planets are equally
likely to orbit the primary and companion star, the median correction factor
became 1.85.  
\citet{hirsch17} analyzed those companions from \citet{furlan17a} found
within 2\arcsec\ of the primary star and with photometric measurements in 
at least two filters. They performed isochrone fits to estimate the stellar
parameters of the companion stars and determined whether the detected
companions are likely to be bound. Confirming the results of \citet{horch14},
they found that most sub-arcsecond binaries are bound; about half of all 
companions at 2\arcsec\ are bound. Using their results from the isochrone fits, 
\citet{hirsch17} derived an average planet radius correction factor of 1.65, 
assuming equal likelihood for the primary and secondary star to be hosting 
the planets.

The effect of changing the planet radius on its density is shown in Figure 
\ref{density_radius}. A correction factor of 1.5 for the planet radius translates 
to a factor of 3.4 decrease in density. We note that, while average correction 
factors for planet radii give an idea of the overall expected changes in planet 
radii, each individual planet will have an individual planet radius correction 
factor depending on its stellar system's configuration and which star the
planet orbits. 
If a stellar system consists of two equal-brightness stars with the same stellar 
radii, the radius of the planet (derived assuming the star is single) would have 
to be revised upward by a factor of $\sqrt 2$, resulting in a decrease in density 
by a factor of 2.8. If the primary star is brighter than the secondary star and the 
planet orbits the primary star, the correction factors for the radii are smaller and 
thus the density decreases less. However, if a star has a relatively faint companion 
and the planet actually orbits this faint star, the radius of the planet can change by 
a factor of a few, and thus the density could decrease by 1-2 orders of magnitude. 

\begin{figure*}[!t]
\centering
\includegraphics[scale=0.65]{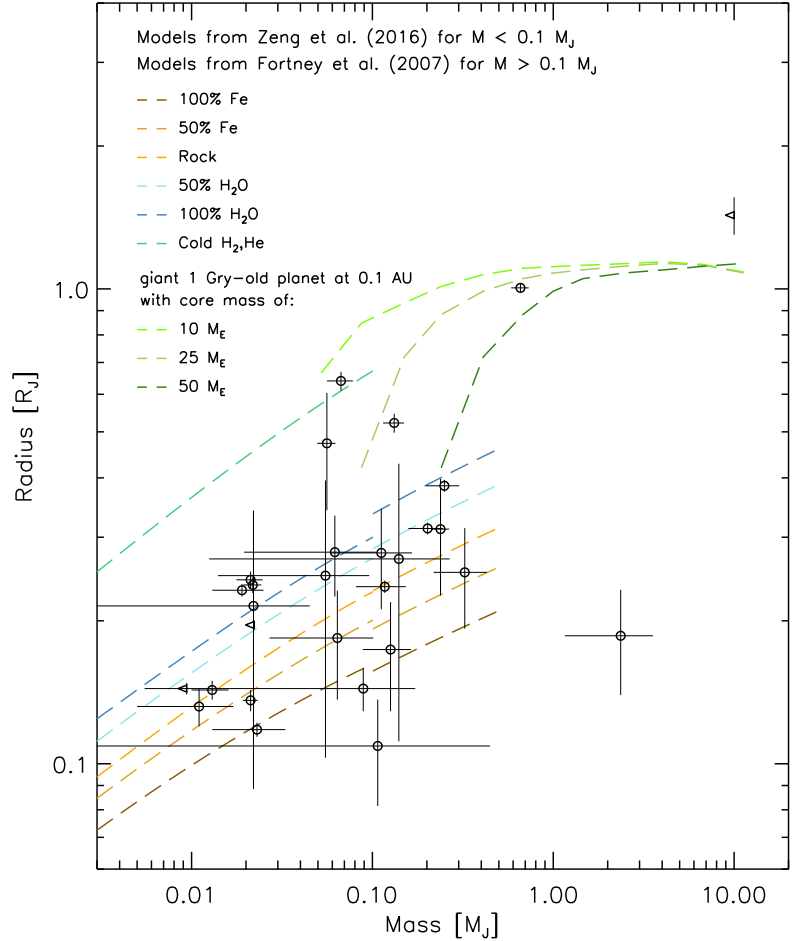}
\caption{Radius versus mass for confirmed {\it Kepler} planets whose host stars 
have stellar companions within 2\arcsec\ and which still require corrections to 
their radii. The colored, dashed lines represent planet models with different 
interior composition from \citet{zeng16} for $M < 0.1\,M_J$ and from
\citet{fortney07} for $M > 0.1\,M_J$ (see label).
\label{KOI_planets_M_R}}
\end{figure*}

\begin{figure*}[!t]
\centering
\includegraphics[scale=0.54]{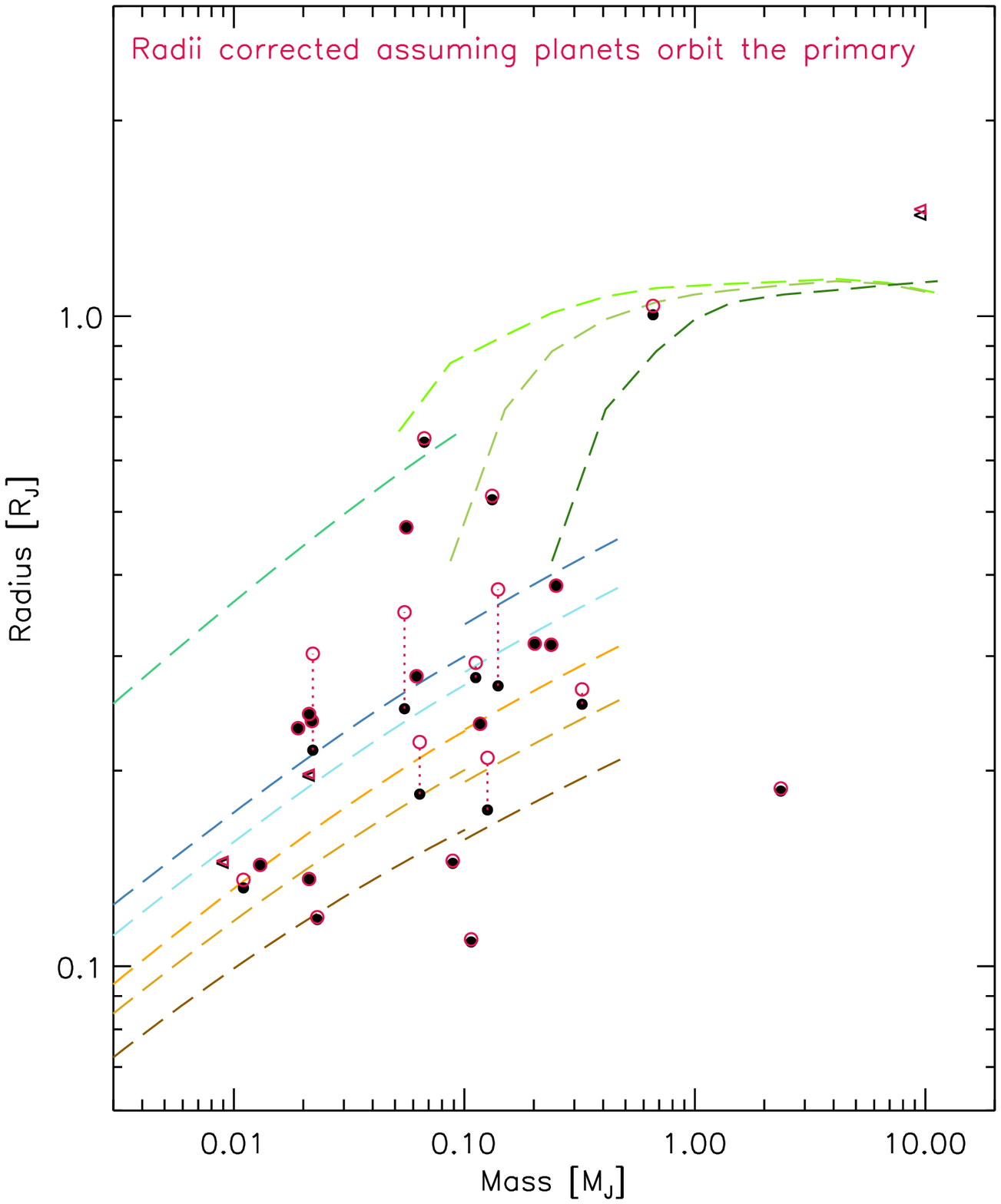}
\includegraphics[scale=0.54]{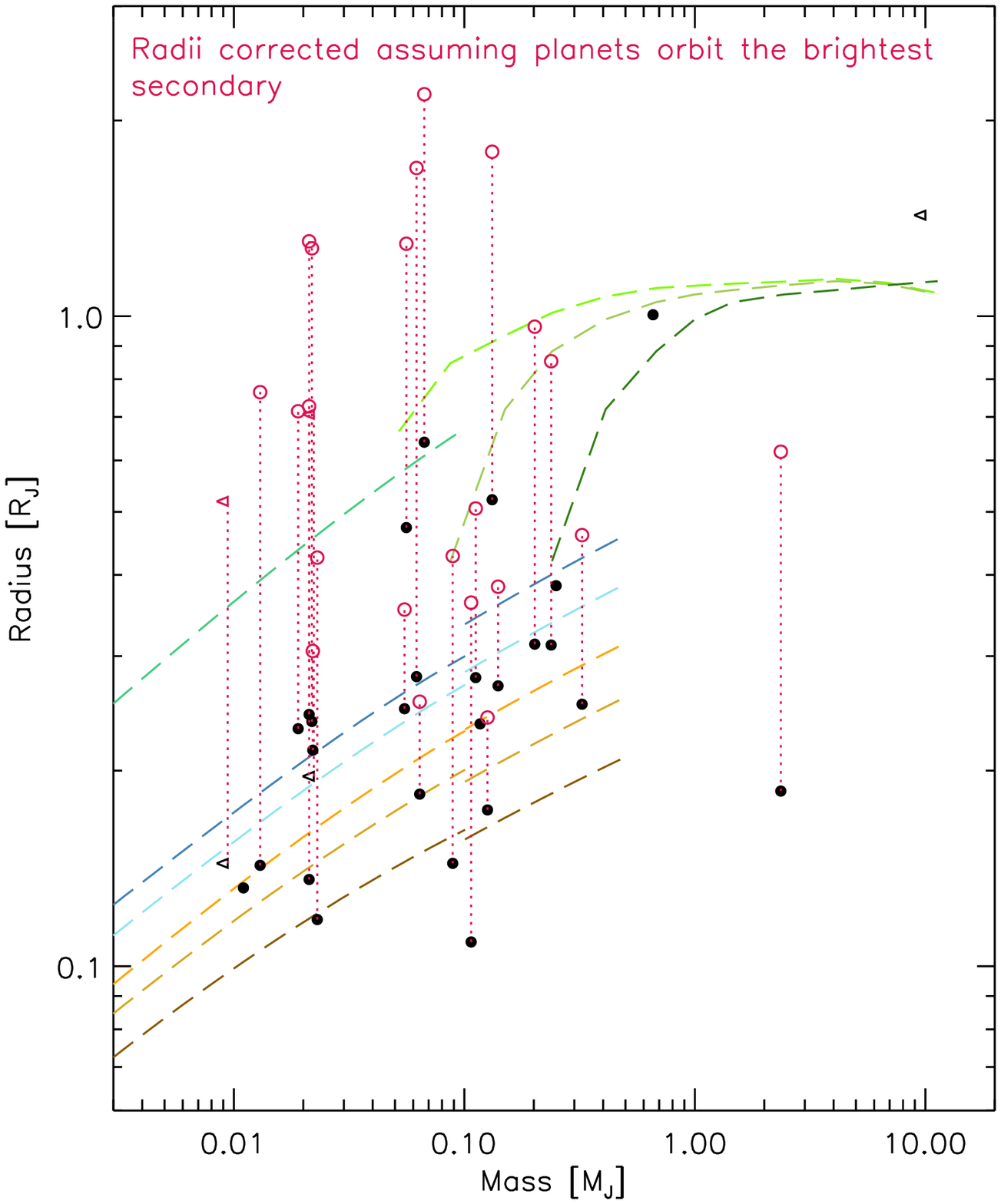}
\caption{Similar to Figure \ref{KOI_planets_M_R}; the black circles are measurements,
(for clarity, uncertainties are omitted), while the red circles result from correcting the 
radii assuming the planets orbit the primary ({\it left}) or brightest companion star 
({\it right}). The colored, dashed lines have the same meaning as in Figure
 \ref{KOI_planets_M_R}.
\label{KOI_planets_M_R_corr}}
\end{figure*}

\subsection{Planet Density and Composition}
\label{densities}

In Figure \ref{KOI_planets_M_R} we plot the radii versus the masses of the
{\it Kepler} planets from Table \ref{planet_densities} (masses and radii
are listed in Table \ref{planet_overview}).
Also shown are model-derived mass--radius relations from \citet{fortney07} 
and \citet{zeng16}; these models allow us to estimate the bulk composition of 
the planets in our sample and to evaluate how the densities change when the 
radii are corrected due to the presence of a stellar companion. For planets with 
masses in the $\sim$ 0.005--0.5 $M_J$ range (which corresponds to 1.6 to 160 
\ME), the composition becomes more volatile-rich the larger the planet radius is; 
for example, with a mass of 0.01 $M_J$ ($=$ 3.2 \ME), a planet with a radius of 
0.1 $R_J$ ($=$ 1.1 \RE) is expected to be composed of pure iron, while a radius 
larger by 30\% and 70\% implies a rocky and 100\% water composition, respectively. 
To infer that this planet has an extensive hydrogen-helium atmosphere, its original 
radius of 0.1 $R_J$ would have to be larger by a factor of 3.6, or equal to
4.0 \RE. Planets with masses larger than about 0.1 $M_J$ are expected 
to have inflated atmospheres if their radii are larger than $\sim$ 1.1 $R_J$ 
\citep{lopez14}.

Figure \ref{KOI_planets_M_R_corr} shows the same data points as Figure
\ref{KOI_planets_M_R}, but for each planet, two points are shown: one with the 
originally derived radius, and one with the radius corrected using the planet 
radius correction factors from \citet{furlan17a}. The left panel of the figure
shows radii corrected with the primary factors, while the right panel displays radii
corrected with the secondary factors. Since for these {\it Kepler} systems 
at least one companion star is present, {\it even if planets orbit their primary star}, 
a correction to the radius is needed.
For those planets found to orbit the primary star (Kepler-74 b, Kepler-97 b,
Kepler-145 b and c, and Kepler-448 b), no corrected radius is shown in the right 
panel of Fig.\ \ref{KOI_planets_M_R_corr}. As mentioned in section 
\ref{sample}, even though there are RV measurements for Kepler-100, none of 
the planets in that system has a clear RV signal detection, and therefore we 
include them in both panels of Fig.\ \ref{KOI_planets_M_R_corr}.

\begin{figure*}[!t]
\centering
\includegraphics[scale=0.6]{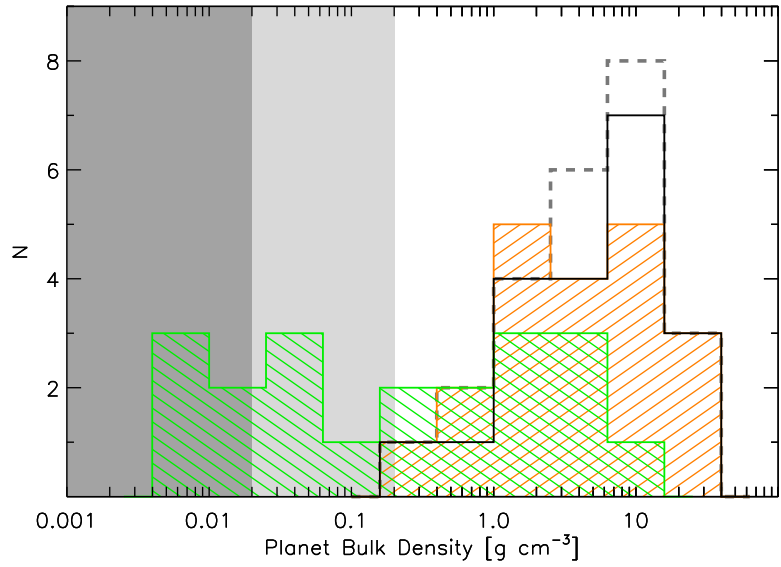}
\caption{Histograms of the bulk densities of confirmed {\it Kepler} planets studied
in this work (see Table \ref{planet_densities}). The gray dashed histogram 
shows all the measurements, excluding upper limits, while the black histogram 
shows only those planets for which both the primary and secondary density correction 
factor is defined (see text for details). The orange and green histograms show the 
densities after correcting the planet radii assuming the planets orbit the primary 
or brightest companion star, respectively. The dark gray area covers unphysically 
low densities, while the lighter gray area covers densities of highly inflated planets.
\label{KOI_planets_rho}}
\end{figure*}

\begin{figure*}[!t]
\centering
\includegraphics[scale=0.62]{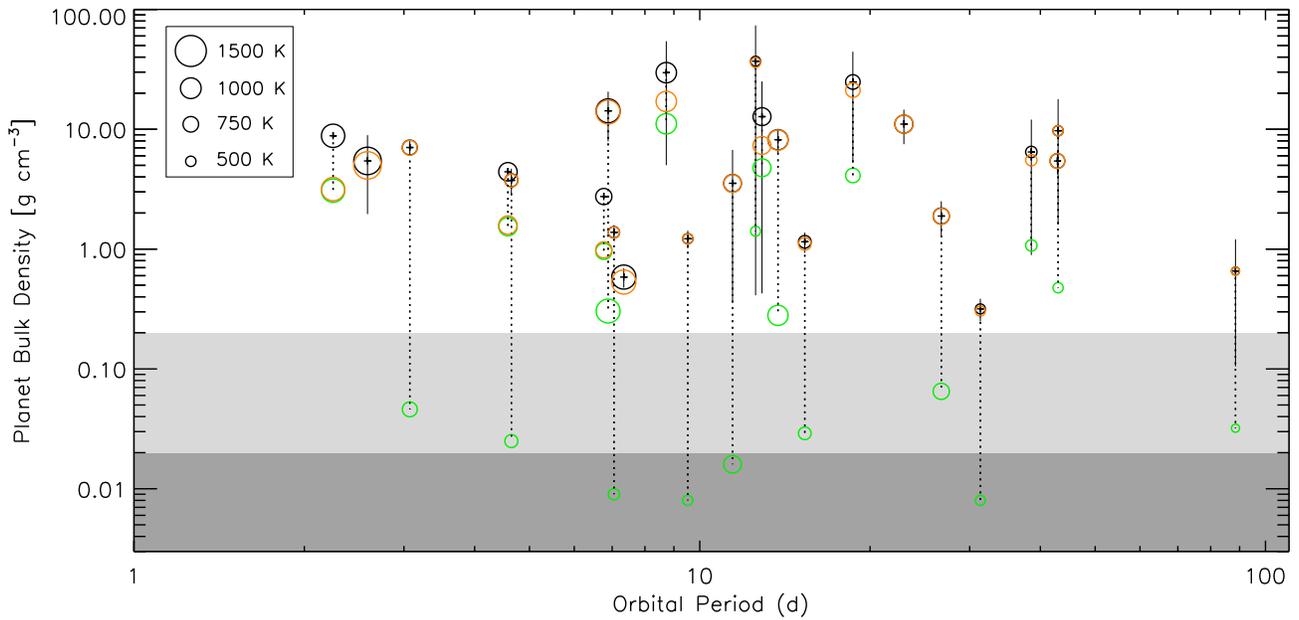}
\caption{Bulk densities from Table \ref{planet_densities} versus the planet
orbital period; the symbol sizes scale with the planet's equilibrium temperature
as shown in the label. Black circles represent density measurements (and vertical, 
solid lines their uncertainties), while the orange and green circles represent 
densities after correcting the planet radii assuming the planets orbit the primary 
or brightest companion star, respectively. The gray areas have the same meaning 
as in Figure \ref{KOI_planets_rho}.
\label{KOI_planets_rho_P}}
\end{figure*}

The planets shown in Figure \ref{KOI_planets_M_R} span a variety of bulk
compositions, from iron-rich, volatile-free planets to more water-rich ones and
planets with extensive atmospheres. Many planet masses (and, to a lesser extent,
planet radii) are very uncertain, and so there is a range in possible planet composition.
In Figure \ref{KOI_planets_rho} we show histograms of the planet bulk densities,
both for measured values and for values corrected due to the presence of a 
companion star using the PDCFs from Table \ref{planet_densities}. The measured 
values range from 0.32 g cm$^{-3}$ to over 20 g cm$^{-3}$ (with the 
latter values very uncertain; see Table \ref{planet_densities}). 
Figure \ref{KOI_planets_rho_P} displays the same bulk densities from Figure 
\ref{KOI_planets_rho} as a function of orbital period, with symbol sizes scaled 
according to the planet's equilibrium temperature (which was adopted as either 
an average of published values, if available, or as the value from the Q1-Q17 
Data Release 25 KOI table). It is expected that planets with short orbital 
periods are hot and, if they have extensive atmospheres, they may be inflated 
and thus have low densities. Indeed, about 40\% of the planets in our 
sample with periods less than 10 days have equilibrium temperatures larger 
than 1000 K, while the planets with longer periods ($>$10 d) are all cooler 
than 1000 K.

When correcting the planet radii due to the flux dilution by the companion
star, for 22 of the 29 planets in our sample the radii and thus also 
densities do not change noticeably if the planets are assumed to orbit their 
primary stars. Only three stars have companions that are bright enough to cause an
obvious increase in the planet radius when accounting for its flux dilution. 
Kepler-326 and Kepler-84 are almost equal-brightness binaries, and so the radii 
of Kepler-326 b, c, d, and of Kepler-84 b and c increase by factors of 1.4 and 
1.2, respectively. The most dramatic change occurs for the Kepler-326
planets, which, with their larger radii, are dominated by gaseous atmospheres
(as opposed to a rock-volatiles mixture before radius correction). The two 
planets in the Kepler-53 system experience a 5\% change in radius.
   
On the other hand, the changes can be substantial if planets are assumed to 
orbit the brightest companion star. In the latter case, most planets whose 
current density identifies them as rocky or water-rich would become gas giants.
However, a large fraction of these planets would reach unrealistically low 
densities ($\lesssim$ 0.1 g cm$^{-3}$), which would require highly inflated atmospheres
(and high equilibrium temperatures) or unusually large ($\gtrsim$ 10\%) 
mass fractions in a H/He envelope, both of which would not be stable, long-lived 
configurations \citep{lopez12}. 
Currently, the planets with the lowest densities (0.02-0.05 g cm$^{-3}$) are 
K2-97 b, Kepler-51 b,c,d, and HAT-P-67 b; K2-97 b and HAT-P-67 b orbit evolved 
stars and have highly inflated atmospheres \citep{grunblatt16,zhou17}, while the 
three planets in the Kepler-51 system either have massive H/He envelopes, or 
their masses are underestimated, given that they were determined via TTVs 
\citep{masuda14}. In Figures \ref{KOI_planets_rho} and \ref{KOI_planets_rho_P},
the region of very low-density planets ($\sim$ 0.02-0.2 g cm$^{-3}$) is indicated by
a light gray area, while densities lower than that (which are improbable, and thus
likely unphysical) are encompassed by a dark gray area. For those planets that 
end up in the low-density regime ($\lesssim$ 0.1 g cm$^{-3}$) after radius correction 
(8 of the 22 planets that could potentially orbit the companion star), 
the scenario of the planet orbiting the companion star can be excluded with a high 
degree of certainty. This includes Kepler-396 c; even though the density of Kepler-396 b 
would allow it to orbit the companion star, Kepler-396 c makes it unlikely for both 
planets to orbit the companion star. Overall, based on their masses, radii, and flux 
contamination by the companion star, we find that 15 planets in 7 planetary 
systems could orbit either the primary or companion star (Kepler-53 b and c, 
Kepler-84 b and c, Kepler-92 b and c, Kepler-100 b, c, d, Kepler-203 c and d, 
Kepler-326 b, c, d, and Kepler-333 b).
   
\clearpage

\section{Discussion}
\label{discuss}

The density of a planet depends on both its mass and radius. While 
different methods exist to determine a planet's mass, some with fairly
large uncertainties, the radii of transiting planets are usually known with
smaller uncertainties than the mass (see Figure \ref{KOI_planets_M_R}).
However, the fact that many stars have nearby companion stars adds
additional uncertainty to the radius determination. When companion stars
have been detected in high-resolution imaging or spectroscopic follow-up
observations, corrections to the planet radii due to flux dilution can be applied;
they are usually relatively small if planets are assumed to the orbit the primary 
star, but can be large if planets orbit the companion star. Of particular concern 
are close binaries of about equal brightness (possibly $\sim$ 15\% of stars); they 
require the largest correction in radius and thus density for planets orbiting 
the primary star (factors of $\sim$ 1.4 and 0.35, respectively). In our sample,
Kepler-326 and Kepler-84 have such a bright, close companion and therefore
experience the most significant change in the bulk composition of their planets.

In most cases, it is not known which star the planet orbits. Besides for
very faint companion stars, which would result in planet radii larger than that
of the star, RV measurements can allow us to exclude a companion star as 
the host, since the primary star's spectrum is the source of the RV information
from which the planet mass is derived. However, for equal-mass (and thus 
equal-brightness) binaries, it could be difficult to distinguish which star is indeed
the planet host; on the other hand, in this case the radius correction factors are
similar for both stars in the system.  
In several multi-planet systems, planet masses have been determined from 
TTVs; in these cases, as opposed to RV detections, the star hosting the planets
is not obvious. The only fairly certain assertion for systems with more than one
planet is that all planets likely orbit the same star.

The planet bulk density can offer an important clue as to whether a planet
can indeed orbit a companion star, given that in this case the density can decrease 
substantially (1-2 order of magnitude).
Low-density planets are known; many can be found in compact, multi-planet
systems (e.g., Kepler-11, \citealt{lissauer13}; Kepler-51, \citealt{masuda14}; 
Kepler-79, \citealt{jontof-hutter14}). Among {\it Kepler} planets, Kepler-51
b, c, d, and Kepler-79 d have the lowest densities measured to date, ranging
from 0.03 g cm$^{-3}$ to 0.09 g cm$^{-3}$ \citep{masuda14,jontof-hutter14}.
Assuming their masses are not underestimated, their low density implies 
that their compositions are dominated, either by volume or mass, by volatiles. 
If the incident flux is sufficiently high (a few hundred times the flux the Earth
receives from the Sun), the atmosphere can be highly inflated, also resulting in
a low density \citep{lopez14}.

The accretion of large amounts of volatiles onto a forming planet presents its 
own challenges; according to one model of giant planet formation, a core has 
to form first, and then sufficient amount of gas has to be available to be accreted 
\citep[see][]{helled14}. These conditions can be met beyond the snow line, with 
subsequent type I migration inward \citep[e.g.,][]{rogers11}. Planets with large 
H/He envelopes and relatively small cores ($\sim$10\%-15\% of volume) could 
be young or could have inflated radii due to strong stellar irradiation; these 
atmospheres could also suffer from photoevaporation and thus become less 
massive over time \citep{rogers11, lopez12}. This atmospheric mass loss
depends on the mass and size of the planet, as well as the stellar UV flux; it is 
expected to be strongest during the first few hundred Myr and could lead to the
complete loss of an atmosphere in several Gyr for a planet with a mass of a 
few \ME\ \citep{rogers11,lopez12}. For planets with substantial atmospheres observed 
today, the atmospheric erosion would imply that the H/He envelopes were even 
more massive in the past, which compounds the challenge of forming substantial 
gaseous envelopes when the planet is still embedded in its protoplanetary disk.
Overall, low-density planets seem to require special formation scenarios and 
conditions and are therefore expected to be rare. In turn, this might imply that 
few exoplanets orbit faint companion stars. 
Out of the 22 planets in our sample that could potentially orbit the companion 
star, we conclude that 8 can only orbit the primary star, since otherwise their 
densities would become lower than $\sim$ 0.1 g cm$^{-3}$.

We note that our sample of 29 {\it Kepler} planets does not include any 
planets comparable to Earth in mass and size. Among the larger sample of
{\it Kepler} planets with masses, radii, and companion stars within 2\arcsec,
the planet with the smallest mass, Kepler-11 f, has a mass of 2.1 \ME\ and a 
radius of 2.5 \RE, while the two planets with radii less than 1 \RE, Kepler-106 b 
and d, only have upper limits in their masses ($<$ 5.4 and 7.9 \ME, respectively). 
The effect of stellar companions on planet radii will be even more important for 
small, presumably rocky planets, since lower densities will imply more volatiles 
and possibly large atmospheres, conditions that are not suitable for life as we 
know on Earth. One problem with Earth-sized planets is that their masses 
are difficult to measure; in many cases, only radii will be measured directly. 
If mass--radius relationships are to be used to infer their masses (and densities), 
it is crucial to determine their radii accurately, which implies detecting any 
nearby companion star.

\section{Conclusions}
\label{conclude}

Given that about half the stars in the solar neighborhood are in multiple
systems, and moreover about 15\% of them have a close, roughly equal-mass
companion, it is important to determine whether a planet host star has
a companion star. The presence of a companion will have an effect on the
determination of the radius of a transiting planet due to the dilution of the
transit depth. 
We studied the effect of companion stars on the radii, and thus bulk
densities, of those confirmed {\it Kepler} planets that have both masses and
radii determined and whose stars have at least one stellar companion detected
within 2\arcsec\ that has not yet been taken into account when deriving
the planets' radii and that could, in most cases, potentially be the planet host.
Our sample contains 29 planets orbiting 15 stars.
In a multiple star system, it is often not known which star the planets orbit, 
but in either case the planetary radii will have to be revised upward. 
Even if the assumption is made that the planets are more likely to 
orbit the primary star, the planet radii would require an increase by as much
as a factor of 1.4, and a corresponding decrease in bulk density by as
much as a factor of 2.8. Such a decrease in density would change the 
composition of any iron-rich planet to that of a planet with at least some 
volatiles, and a rocky planet would become a planet dominated by volatiles. 

Even more dramatic changes in the inferred planet bulk composition are 
expected if the planet orbits a fainter companion star; in this case several
planets in our sample would be inferred to have extensive hydrogen/helium 
atmospheres (likely also highly inflated). This scenario is probably not very common, 
and it can be ruled out if the planet bulk density would become unrealistically low, but 
it has to be assessed on a case by case basis. Of particular interest are small, 
rocky planets; they are more affected by the presence of companion stars, 
since they could still be Earth-like (if orbiting the primary star) or dominated 
by volatiles (if orbiting a fainter companion star), and thus not be Earth-like 
at all. Since masses are very challenging to measure for small planets, it is 
critical to at least determine accurate radii for them in order to derive a good 
estimate of their mean density.
 
Of the 29 planets in our sample, seven experience notable increases 
in their radius once the effect of the companion star is folded in: Kepler-326 
b, c, d, Kepler-84 b and c, and, to a lesser extent, Kepler-53 b and c. 
In particular, the Kepler-326 planets would change from a composition of 
rock and some volatiles to one dominated by a gaseous envelope. 
Five planets in our sample cannot orbit the companion star, since 
previous work determined that they orbit the primary star. Of the 
remaining planets with measured densities, eight would end up with 
unrealistically low densities if they orbited the companion star. 
Overall, we conclude that in seven planetary systems (with a total 
of 15 planets) the planets could orbit either the primary or the companion 
star (Kepler-53, Kepler-84, Kepler-92, Kepler-100, Kepler-203, Kepler-326, 
and Kepler-333).

The effect of a companion star on the bulk density of a planet underlines the 
importance of follow-up studies of host stars of planet candidates found 
with the transit method. High-resolution imaging and radial velocity measurements
will reveal companion stars in certain ranges of parameter space; in addition, 
in-depth statistical analysis using the observational results should allow us to 
infer which star the planet is most likely to orbit. Among the {\it Kepler} planet 
host stars, there are likely still many unidentified binary systems; for host stars 
that are closer (and brighter), as is the case for many K2 and most TESS targets, 
fewer companions are missed by follow-up observations \citep[e.g.,][]{vanderburg15,
crossfield15, ciardi15, howell16}. Thus, with appropriate follow-up work, the large 
expected planet yield of the K2 and TESS missions, as well as other future 
transiting surveys, should result in more reliable planet radii and therefore more 
definitive identification of truly Earth-like planets in the solar neighborhood. 

\acknowledgments
We thank our referee for useful suggestions that improved the 
presentation and clarity of the paper.
Support for this work was provided by NASA through awards issued by
JPL/Caltech.
This research has made use of the NASA Exoplanet Archive and the Exoplanet
Follow-up Observation Program website, which are operated by the California 
Institute of Technology, under contract with NASA under the Exoplanet Exploration 
Program.
It has also made use of NASA's Astrophysics Data System Bibliographic Services.

\appendix

\section{Notes on Individual Planets}

\subsection{Planets Studied in This Work (Targets from Table \ref{planet_densities})}

{\it Kepler-27 b and c. ---}
The masses of Kepler-27 b and c were determined from TTVs by \citet{hadden14};
since these authors identified the planet pair as likely having high eccentricities, their
masses are probably overestimated. From the measured masses and radii, we
derive bulk densities of 1.15 and 0.32 g cm$^{-3}$ for Kepler-27 b and c, respectively.
Both planets seem to be gas giants, with a smaller core for Kepler-27 c than b.
\citet{steffen12} raised the possibility that the planets could actually orbit the 
companion star $\sim$ 2\arcsec\ to the northeast ($\Delta K=$ 3.4; \citealt{furlan17a}).
However, for this scenario we derived unrealistically low planet densities 
($<$ 0.03 g cm$^{-3}$), which would decrease even more if the planet masses
were actually lower. 

{\it Kepler-53 b and c. ---}
Similar to the planets of Kepler-27, the masses of Kepler-53 b and c were determined
from TTVs, and they are identified as high-eccentricity planets \citep{hadden14}. We
derive high bulk densities for both planets, but the uncertainties are large. Kepler-53 b
is consistent with an iron-rich rocky composition, while Kepler-53 c is water-rich.
If the dilution caused by the 0.1\arcsec\ companion ($\Delta$m $\sim$ 2.5 at 0.55 
$\mu$m; \citealt{gilliland15}) is taken into account, the planet densities decrease by 
about 15\% if the planets are assumed to orbit the primary star, but by almost 85\% if
they are assumed to orbit the fainter star. In the latter case, both planets would be
inferred to have substantial gaseous envelopes.

{\it Kepler-74 b. ---}
Kepler-74 b is a gas giant planet whose mass was determined from RV measurements
\citep{hebrard13,bonomo15}. It has a relatively high equilibrium temperature of $\sim$
1200 K. The star has a companion at a separation of 0.3\arcsec\ which is about 0.5 mag
fainter in the optical \citep{ziegler17}. This companion was not taken into account when
the planet radius was derived, and so, even though the planet is orbiting the primary
(given its RV signal), its radius has to be revised by about 3\%.

{\it Kepler-80 b to f. ---}
Kepler-80 is surrounded by five transiting planets \citep{xie13,lissauer14,rowe14,morton16},
all of which have orbital periods less than 10 days. The densities of the four planets with 
mass and radius determinations imply a rocky composition for Kepler-80 d and e and 
substantial atmospheres for planets b and c \citep{macdonald16}. We can exclude the 
scenario that the planets orbit the 1.7\arcsec-companion star of Kepler-80 ($\Delta K=$5.2; 
\citealt{kraus16}) since the densities of at least some of the planets in each system would 
become unrealistically low.

{\it Kepler-84 b and c. ---}  
Kepler-84 is among those systems for which the planet radius, and thus planet 
density, changes substantially even if the planets orbit their primary star, since 
the companion star at 0.2\arcsec\ is only somewhat fainter than the primary 
($\Delta$m $\sim$ 0.9 at 0.55 $\mu$m; \citealt{gilliland15}). Its primary and 
secondary planet radius correction factors are 1.202 and 1.387, respectively 
\citep{furlan17a}. It is surrounded by five planets, of which only two have 
measured masses from TTVs \citep{hadden14}. The composition of Kepler-84 
b and c implies iron-rich solids; with the larger planet radii, they would still be 
rocky planets, but in the case of Kepler-84 c (whose mass is about a factor of 
two lower than that of Kepler-84 b), the new density suggests the additional 
presence of some water or other volatiles.

{\it Kepler-92 b and c. ---}
The masses of Kepler-92 b and c were determined from TTVs \citep{xie14}. 
With a mass of 0.2 $M_J$, Kepler-92 b is ten times as massive as Kepler-92 c.
Their radii both lie in the Neptune-size regime. From their position in the 
mass--radius diagram, their composition is likely rich in volatiles. If they 
orbited the faint companion star instead of the primary (which is about 
4.2 mag fainter in the $K$-band and at a projected separation of 1.5\arcsec;
\citealt{kraus16,furlan17a}), Kepler-92 c would be among the among the 
lowest-density planets, while Kepler-92 b would be a typical gas giant.

{\it Kepler-97 b. ---}
Kepler-97 is orbited by two planets, but only one (Kepler-97 b) has both mass and 
radius determined; the other one, Kepler-97 c, was detected in radial velocity 
data as a linear trend, so only a lower limit of $\sim$ 1 $M_J$ for its mass 
could be derived \citep{marcy14}. Kepler-97 has a fainter companion at a 
separation of 0.4\arcsec\ ($\Delta K=$3; \citealt{furlan17a}); \citet{marcy14} 
suggested that it could be the cause for the linear trend in the RVs (and thus 
Kepler-97 c would not exist). They also concluded that Kepler-97 b most likely 
orbits the primary star due to to the lack of centroid shift when comparing the 
in- and out-of-transit photocenter. However, \citet{marcy14} did not correct
the planet radius due to the flux dilution caused by the companion star (since
it is a relatively small correction of 3\%; \citealt{furlan17a}).

{\it Kepler-100 b, c, d. ---}
There are three planets in the Kepler-100 system, but only one, Kepler-100 b, 
has a measured mass from RV data, albeit with just a tentative RV signal detection
\citep{marcy14}. Given the tentative RV signal of just one planet, we did not exclude 
the fainter companion star (at a projected separation of 1.8\arcsec, with $\Delta i=$4.2; 
\citealt{lillo-box14,furlan17a}) to be the planet host. The density of Kepler-100 b 
implies an iron-rich composition, but since its mass is fairly uncertain, it could be 
more rich in volatiles. The upper limits in mass for Kepler-100 c and d suggest 
they are volatile-rich planets. If the planets orbited the companion star, they 
would be low-density giant planets, with Kepler-100 c and d among the 
lowest-density planets known. 

{\it Kepler-104 b. ---}
The Kepler-104 multi-planet system contains three planets of similar size
\citep{rowe14}, but only Kepler-104 b has a mass determined from TTVs 
\citep{hadden14}. Since it is flagged as a high-eccentricity planet by 
\citet{hadden14}, its mass could be overestimated. Its measured mass implies 
a composition dominated by volatiles. If Kepler-104 b orbited the faint companion 
star at 1.9\arcsec\ from the primary ($\Delta i=$6.1; \citealt{lillo-box14}), its density 
would become unrealistically low.

{\it Kepler-145 b and c. ---}
The masses of Kepler-145 b and c were determined from TTVs \citep{xie14}. 
Kepler-145 c is both larger and more massive than Kepler-145 b; its composition
is likely dominated by volatiles, while Kepler-145 b is mostly rocky. The very
faint companion star at a projected separation of 1.5\arcsec\ ($\Delta K=$8.5; 
\citealt{kraus16}) causes a negligible flux dilution (primary planet radius correction
of 1.0001; \citealt{furlan17a}); we conclude that the companion cannot be the planet
host since the planets would become bigger than the star.

{\it Kepler-203 c and d. ---} 
Kepler-203 is orbited by three planets \citep{rowe14}, but only the two outermost
planets, Kepler 203-c and d, have masses determined from TTVs \citep{hadden14}.
The location of Kepler-203 c and d in the mass--radius diagram implies a density
higher than pure iron. However, they were flagged as high-eccentricity planets, 
which suggests that their masses are overestimated \citep{hadden14}. Masses lower 
by at least an order of magnitude would make these planets consistent with a rocky 
or water-rich composition.  On the other hand, if Kepler-203 c and d transited the 
companion star (located at 1.9\arcsec, with $\Delta i$=4.1; \citealt{lillo-box14,furlan17a}) 
instead of the primary, their radii would be larger by about a factor of three, and thus, 
even if their masses did not change, their density would be low enough to be consistent 
with that of a gas giant planet.

{\it Kepler-326 b, c, d. ---}
Similar to Kepler-84, the planets of Kepler-326 require substantial revisions
to their radii and densities even if they orbit the primary star.
The primary and secondary planet radius correction factors are 1.407 and 1.421, 
respectively (resulting from an almost equal-brightness binary; the two stars are
just 0.05\arcsec\ apart; \citealt{kraus16}). Kepler-326 has three planets, all of which 
have measured masses from TTVs \citep{hadden14}. After correcting for the flux 
dilution by the companion, the increased planetary radii imply substantial atmospheres 
as opposed to water-dominated (Kepler-326 c and d) or water-rock (Kepler-326 b) 
composition.

{\it Kepler-333 b. ---}
The mass of Kepler-333 b was determined from TTVs \citep{hadden14}. There
is another planet in the system, Kepler-333 c, without a mass determination, but a 
somewhat smaller radius \citep{rowe14}. The density of Kepler-333 b implies a 
density higher than pure iron. If its mass were lower by at least a factor of 10, its 
composition would be consistent with that of rock or a rock-water mixture. Similar 
to the Kepler-203 system, if Kepler-333 b orbited the companion star (separated by
1.3\arcsec\ from the primary, with $\Delta i=$4.1; \citealt{ziegler17}), its mass and 
radius would be consistent with that of a gas giant planet.

{\it Kepler-396 b and c. ---}
The masses and radii of Kepler-396 b and c imply a volatile-rich composition for
for planet b, while planet c, which is larger and less massive, is similar to a gas 
giant. Since masses were determined from TTVs \citep{xie14}, they may be
overestimated. If the planets orbited the faint companion, separated by 0.6\arcsec\
from the primary and about 6 mag fainter in the optical \citep{furlan17a}, Kepler-396 b
would become an envelope-dominated planet, while the density of Kepler-396 c would
become unrealistically low. Therefore, it is likely that both planets orbit the primary star.

{\it Kepler-448 b. ---}
Kepler-448 b is a 1.4-$R_J$ planet with only an upper limit of 10 $M_J$ for its mass
derived from RV measurements \citep{bourrier15}. Its large radius (1.4 $R_J$) implies
a highly inflated atmosphere (irrespective of the mass of the planet), but, as opposed 
to Kepler-13 b, which has similarly large radius, it does not have a particularly high 
equilibrium temperature. From the analysis of time-series spectra, \citet{bourrier15}
found that the transit is associated with Kepler-448; even though they were not aware
of any companion stars, we conclude that the companion star located at 0.6\arcsec\ from 
the primary (with $\Delta K=$3.8; \citealt{kraus16}) is unlikely to be the planet host.
However, even if Kepler-448 b orbits the primary star, its radius has to be corrected 
due to the flux dilution by the companion (a small increase of 2\%; \citealt{furlan17a}).

\subsection{Remaining Targets from Table \ref{planet_overview}}

{\it Kepler-1 b. ---}
Kepler-1 b, also known as TrES-2 b, was discovered by \citet{odonovan06}; since its
discovery, many authors have measured its properties from the {\it Kepler} light
curve and ancillary data (see Table \ref{planet_overview}). Several authors did correct
for the flux dilution by the faint companion (at 1.1\arcsec\ with $\Delta i \sim$ 4; 
\citep{law14}), but the correction to the planet radius is just $\sim$ 1\% \citep{furlan17a}.
With a mass of 1.22 $M_J$, a radius of 1.21 $R_J$, and an equilibrium temperature 
of 1470 K \citep{esteves15}, Kepler-1 b is a hot Jupiter with a somewhat inflated radius. 
Given that its mass was determined from RV measurements \citep[e.g.,][]{odonovan06}, 
we can exclude the companion star as being the host.

{\it Kepler-5 b. ---}
Kepler-5 b is about a third larger than Jupiter and twice as massive. Its large radius, 
as well as its equilibrium temperature of 1750 K \citep{esteves15}, suggest that its 
atmosphere is inflated. Similar to Kepler-1 b, the mass of Kepler-5 b was determined 
from RV measurements \citep{koch10}, and the companion at 0.9\arcsec\ is about 
5 magnitudes fainter than the primary in the $J$-band \citep{furlan17a}, so the 
companion star cannot be the planet host. Also, the (very small) flux dilution 
by the companion was taken into account when the planet parameters for
Kepler-5 b were derived \citep{koch10,southworth11}.

{\it Kepler-7 b. ---}
Kepler-7b stands out as a planet with a very large radius (1.6 $R_J$), small mass 
(0.4 $M_J$), and thus extremely low density (0.14 g cm$^{-3}$; \citealt{esteves15}). 
its atmosphere is likely highly inflated, a result of its close orbit around a slightly 
evolved star \citep{latham10}, resulting in a high equilibrium temperature of 1630 K
\citep{esteves15}. The host star has a faint companion ($\Delta i=$4.6) at a projected 
separation of 1.9\arcsec\ \citep{latham10,adams12,law14}, which was taken into
account when the planet parameters for Kepler-7b were derived \citep{latham10,
southworth10,southworth11}. Moreover, the companion was excluded as being the 
star being transited due to the small observed centroid shifts, in addition to 
the primary's measured radial velocity shifts \citep{latham10}. 
 
{\it Kepler-10 b and c. ---}
There are two planets in the the Kepler-10 system: both Kepler-10 b and c have a relatively
high density of $\sim$ 7-8 g cm$^{-3}$ \citep{dumusque14,esteves15}, but the former 
has a mass of about 4 \ME\ and a radius of 1.5 \RE, while the latter is more massive and 
larger with 17.2 \ME\ and 2.4 \RE\ \citep{hadden14,lissauer11,lissauer13}. The densities 
of both planets suggests a rocky composition, with Kepler-10 b containing some iron and 
Kepler-10 c likely some water \citep{dumusque14}. The absence of an atmosphere and 
high equilibrium temperature of Kepler-10 b (2130 K; \citealt{esteves15}) suggests that 
it may be the remnant core of a gas-rich planet whose atmosphere was lost due to 
photoevaporation \citep{lopez14}. Given that the masses of the planets in the Kepler-10 
system were derived from RV measurements, the 2\arcsec\ companion star ($\Delta K=$6.8; 
\citealt{kraus16}) cannot be the planet host. Its flux dilution has not been taken into account 
previously, but the correction factor for the planet radius is negligible at 1.0005 
\citep{furlan17a}.

{\it Kepler-11 b to g. ---}
The Kepler-11 system consists of six transiting planets. The masses of five planets
(b through f) were determined from TTVs (planet g has only an upper limit in mass),
and their densities imply large volume fractions of volatiles, like H$_2$O, CH$_4$, 
H$_2$, and He \citep{lissauer11, lissauer13, hadden14}. The planets around Kepler-11
form a tight planetary system; they all orbit within 0.5 au from their star \citep{lissauer13}. 
A faint companion star ($\Delta K=$4.6) lies at separation of 1.3\arcsec\ \citep{wang15a,
kraus16}); its flux dilution has not been considered previously, but it is very small
causing a radius change by less than 0.5\% \citep{furlan17a}. Based on centroid 
analysis of {\it Kepler} data, \citet{lissauer11} concluded that the companion star
cannot be the planet host.

{\it Kepler-13 b. ---}
Kepler-13 b has a very large mass (9.0 $M_J$) and also large radius (1.5 $R_J$),
implying a substantial atmosphere \citep{shporer14,esteves15}. It also has a high
equilibrium temperature of 2550 K \citep{esteves15}. In fact, it was found to likely 
orbit the brighter star of a 1\arcsec-binary, of which both components are rapidly 
rotating A-type stars \citep{szabo11}. With a period of just 1.7 days, it is a hot 
Jupiter with a highly inflated atmosphere \citep[e.g.,][]{shporer14}. The transit 
depth dilution by the companion was taken into account when the radius of 
Kepler-13b was derived \citep{shporer14,esteves15}.

{\it Kepler-14 b. ---}
Similar to Kepler-13 b, Kepler-14 b has a very large mass (8.1 $M_J$; 
\citealt{buchhave11,southworth12}). It is a bit smaller than Kepler-13 b, which 
implies a higher bulk density. The almost equal-brightness companion star at a
separation of just 0.3\arcsec\ was taken into account in the light curve fit and in
the derivation of the radial velocity \citep{buchhave11}. Moreover, from centroid
analysis of {\it Kepler} data \citet{buchhave11} concluded that the primary star is
the planet host.

{\it Kepler-21 b. ---}
The mass and radius of Kepler-21 b, 5.1 \ME\ and 1.64 \RE, respectively, suggest that
it is likely a rocky planet \citep{lopez-morales16}. Given its high equilibrium temperature of
$\sim$ 2000 K (it is in a 2.8-day orbit around a star with $T_{\mathrm{effect}}=$6300 K), 
\citet{lopez-morales16} suggested that the planet is surrounded by a thick layer of 
molten rock; it could be the left-over core of a giant planet. The mass of Kepler-21 b
has been determined from RV measurements, so we can exclude the faint companion 
($\Delta K=$4; \citealt{kraus16,furlan17a}) at a projected separation of 0.8\arcsec\ 
as the planet host star. \citet{lopez-morales16} were not aware of this companion,
so its flux dilution has not been accounted for, but it is very small \citep{furlan17a}.

{\it Kepler-64 b. ---} 
Kepler-64 b is a circumbinary planet; its host star actually forms a quadruple stellar 
system, with an eclipsing binary and wider binary at a separation of $\sim$ 0.7\arcsec\ 
\citep{schwamb13}. In addition, there is a faint star at a projected separation of 3\arcsec. 
The flux dilution by these companions (about 13\%) was taken into account by 
\citet{schwamb13} when deriving the transit depth and thus planet radius of 
Kepler-64 b. They also derived an upper limit for the mass, which, combined with
the radius measurement, places this planet in the gas giant regime. Based on the
RV measurements and models of \citet{schwamb13}, it is clear that the eclipsing
binary star is the host.
 
{\it Kepler-106 b to e. ---}
Kepler-106 has a total of four planets, with RV yielding mass measurements for 
planets c and e and upper limits for the mass of planets b and d \citep{marcy14}. 
Kepler-106 b and d are probably consistent with an iron-rich or rocky composition. 
On the other hand, Kepler-106 c and e are volatile-rich. Given the very faint companion 
star at 1.7\arcsec\ from the primary ($\Delta K=$8.5; \citealt{kraus16}), the transit 
dilution is minimal (and so it does not matter that it was not taken into account by 
\citealt{marcy14}); moreover, the companion star cannot be the planet host, since, 
in addition to the RV detections of planets c and e in the spectrum of the primary, 
in this case the planets would become bigger than the star.

{\it Kepler-424 b. ---}
There are two planets in the Kepler-424 system: Kepler-424 b is a hot Jupiter that
transits the planet during its 3.3-day orbit, and Kepler-424 c is a $\sim$ 7 $M_J$
on a 223-day orbit which was detected in RV data and does not transit \citep{endl14}. 
Thus, only Kepler-424 b has both mass and radius determined. The close companion
star (at 0.07\arcsec) is faint ($\Delta K=$ 3.7; \citealt{kraus16}) and thus would cause 
a very minor correction to the planet radius and density. Thus, similar to some of the
other planets presented here, it does not really matter that it was not taken into
account previously. It cannot be the planet host, since both planets were detected 
in the RV data of the primary.

{\it Kepler-432 b. ---}
Similar to the Kepler-424 system, Kepler-432 is orbited by a transiting planet 
(Kepler-432 b) and another planet (Kepler-432 c), detected in RV data, that does 
not transit \citep{quinn15}. However, the host star is a red giant star, and the orbit of 
Kepler-432 b is very eccentric \citep{ciceri15,quinn15}. Kepler-432 b is also fairly
massive ($\sim$ 5.2 $M_J$), and its radius of 1.13 $R_J$ places it at the lower end
of the inflated gas giant regime. The flux dilution by the 0.9\arcsec\ companion 
($\Delta K=$5.1; \citealt{kraus16,furlan17a}) was taken into account when analyzing 
the light curve of Kepler-432 b \citep{ciceri15}, but the effect is very small. Given 
the RV data, the companion can be excluded as the planet host.

\end{document}